\documentclass[twocolumn,showpacs,preprintnumbers,amsmath,amssymb]{revtex4}
\usepackage{graphicx}% Include figure files
\usepackage{dcolumn}% Align table columns on decimal point
\begin{document}

\preprint{APS/123-QED}

\title{A Unified Theory  of Chemical Reactions}% Force line breaks with \\

\author{S. Aubry}%
 \email{serge.aubry91@gmail.com}
\affiliation{%
Laboratoire L\'eon Brillouin(CEA-CNRS), CEA Saclay,
91191-Gif-sur-Yvette, France}%
%\affiliation{

\date{\today}% It is always \today, today,

\begin{abstract}
We propose a new and general formalism
for elementary chemical reactions where quantum electronic variables are used as reaction coordinates.  This formalism
is in principle applicable to all kinds of chemical reactions ionic or covalent.
Our theory reveals the existence of an intermediate situation between ionic and covalent which may be almost barrierless and isoenegetic
and which should be of high interest for understanding biochemistry.
\end{abstract}

%Use showkeys class option if keyword
\maketitle

Chemical reactions are primarly changes of electronic states (associated with molecular and environmental reorganization) which appears either in radical ionization (redox) or 
in the forming/breaking of chemical bonds. They can be generally decomposed into sequences of elementary reactions corresponding
to single transitions between two different electronic states for example an electron transfer
(ET) between a Donor and an Acceptor.The rate of chemical reactions often obeys the Arrhenius law which manifest the existence of
an energy barrier between the reactants and the products which has to be overcome by the thermal fluctuations. 
This energy barrier is usually quite large compared to the ambient temperature energy ($\approx 0.026 eV$ at $300 K$).
There are also   chemical reactions which do not obey the Arrhenius law (with a positive energy barrier).
This is the situation for  \textit{free radicals} with unpaired electrons which are generally highly reactive
\cite{free_rad} and generate covalent bonds. 

\begin{figure}
    \centering
   \includegraphics[angle=270,width=0.50 \textwidth]{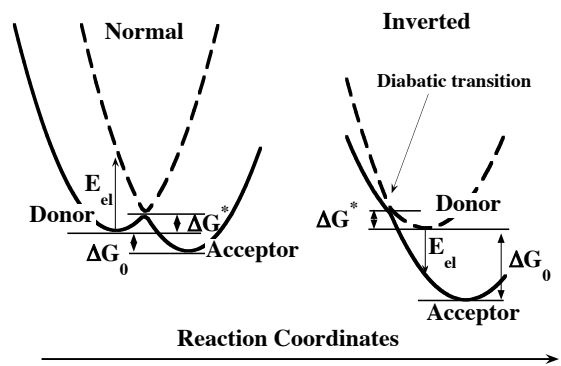}
    \caption{Standard Theory: The surfaces of energy versus Reaction Coordinates are  two intersecting paraboloids, one corresponding to
  the reactants and  one to the products. A small gap opening at intersection determines a lower energy surface
  and an upper energy surface.}
 \label{fig1}
\end{figure}

The standard theory for ET (redox) mostly due to Marcus\cite{Mar93}, considers the free energy the whole system as a function
of the nuclei ( reaction)  coordinates when the electron is on the Donor site (reactants) or  on the Acceptor site (products)
These functions are approximate as paraboloids schematically represented fig.\ref{fig1}.
There are two regimes called normal when at constant coordinates, the electronic excitation from Donor to Acceptor requires 
to absorb a positive energy $E_{el}$ and inverted when $E_{el}$ is negative.

 The lowest point at the intersection  of these two surfaces determines the minimum free energy $\Delta G^{\star}$ to be provided to the system 
 for transferring the electron between Donor and Acceptor.  This energy barrier may be reached because of the thermal fluctuations of the nuclei 
 with a probability per unit time proportional to $e^{-\frac{\Delta G^{\star}}{k_BT}}$ which yields the main factor of the Arrhenius law. At this point 
 the two electronic states on Donor and Acceptor are degenerate so that ET may occur by  quantum tunnelling. Actually, because of small overlap terms.  
 between the electronic orbitals, degeneracy is raised with a small gap so that the two   diabatic surfaces  are  non intersecting as shown  fig.\ref{fig1}. 
Electron hopping from Donor to Acceptor  is considered as a probabilistic event which depends 
on  the time during which resonance lasts and thus depends on the phonon frequencies and the temperature. 
The transition probability between the two diabatic surfaces $A(T)$ which contributes to the prefactor of  the Arrhenius law is empirically  
calculated from  the Landau-Zener model. In the normal regime,  ET does not require any diabatic transition while in the inverted regime  a diabatic transition 
from the upper  to the lower energy surface is necessary (see fig.\ref{fig1}).

A great mystery of the chemistry of life (biochemistry) is that it operates efficiently  about room  temperature
even when involving highly energetic reactions (e.g. oxidation of sugars). Most the released  energy remains stored in other chemical forms 
to be involved later for fueling subsequent bioreactions.  
Thus, biochemistry essentially operates along nearly isoenergetic chemical paths  that is with energy variations and barriers  smaller or comparable to the room temperature energy  ($0.026 eV$ at $300 K$). Enzymes generate such chemical paths by triggering  specific chemical reactions which  could not occur spontaneously. Backward reactions could even be 
also catalysed under changes in the environmental conditions.
The unified theory which we outline now, show that  such chemical paths may exist by involving both charge and covalent interactions in well-tuned systems (as should be the enzymatic systems).

We start from first principles and consider very generally, the global quantum  Hamiltonian $H$ of our reacting system
with interacting electrons $\alpha$ with coordinates $\mathbf{r}= \{\mathbf{r}_{\alpha}\}$
also interacting with a collection of quantum nuclei $i$ with masses $M_i$ and coordinates $\mathbf{R}_i$. It can be written with the general form
$H =  \sum_i \frac{\mathbf{P}_i^2}{2M_i}+H_e(\mathbf{R}) $
as the sum of the kinetic energy operator  of the nuclei which is a function of the conjugate operators $\mathbf{P}_i= \frac{\hbar}{i} \nabla_{\mathbf{R}_i}$ of
the nuclei coordinates $\mathbf{R}_i$ and the Hamiltonian $H_e(\mathbf{R})$ which describes the rest that is the whole system of electrons 
submitted to a potential generated by the nuclei. 
The standard Born-Oppenheimer (BO) approximation assumes that the global wavefunction has the form $\Psi(\mathbf{r},\mathbf{R},t)= \Phi(\mathbf{R},t) \psi_0(\mathbf{r},\mathbf{R})$
where $\psi_0(\mathbf{r},\mathbf{R})$ is the electronic groundstate of Hamiltonian $H_e(\mathbf{R})$. Its eigen energy $E_0(\mathbf{R})$  becomes the effective potential for the nuclei.

We consider now an elementary chemical reaction corresponding to a transition between two electronic states called Donor (initial) and Acceptor (final) so that the electronic state remains confined in a 2D subspace $\mathcal{E}(\mathbf{R})$  invariant by $H_e(\mathbf{R})$ spanned by these two states.  All the other electronic eigenstates (far in energy) are discarded. It is convenient to choose a  LCAO  real and  orthogonal base  $\psi_D(\mathbf{r},\mathbf{R})$ representing for example to the electron on a donor site and   $\psi_A(\mathbf{r},\mathbf{R})$ this electron on an acceptor site. The global wave function has  the form $\Psi(\mathbf{r},\mathbf{R},t)= \Phi_D(\mathbf{R},t) \psi_D(\mathbf{r},\mathbf{R})+ \Phi_A(\mathbf{R},t)\psi_A(\mathbf{r},\mathbf{R})$.

Then, integration over the electronic variables $\mathbf{r}$ yields the nuclei Hamiltonian $<\Psi(\mathbf{r},\mathbf{R},t)|H_e(\mathbf{R})+ \sum_i \frac{\mathbf{P}_i^2}{2M_i}|\Psi(\mathbf{r},\mathbf{R},t)>_r  = \tilde{H}_e+\tilde{K}$ which now operates  in the two-components  wave function space $\Phi(\mathbf{R},t) = \left(\begin{array}{ c}
     \Phi_D(\mathbf{R},t)   \\
     \Phi_A(\mathbf{R},t)  
\end{array} \right)$. $\tilde{H}_e$ is a $2\times 2$ matrix which has the form $\tilde{H}_e= \left(\begin{array}{ cc}
E_D(\mathbf{R})       &   
\Lambda(\mathbf{R})\\
\Lambda(\mathbf{R} )   &  E_A(\mathbf{R})
\end{array} \right) $ only dependant on $\mathbf{R}$ while the projected kinetic energy  operator 
$\tilde{K}=\sum_{i,\alpha} \frac{P_{i,\alpha}^2}{2M_i}$ can  be expressed with the following overlap integrals
$a_{i,\alpha}^{n,m}(\mathbf{R}) =  \frac{1}{M_i}  \int \psi_n(\mathbf{r},\mathbf{R}) \frac{\partial  \psi_m(\mathbf{r},\mathbf{R})} {\partial R_{i,\alpha}}  d\mathbf{r} $ for $n,m=D$ or $A$. 
Orthonormalization implies  $\mathbf{a}^{D,D}(\mathbf{R})=0$ and $\mathbf{a}^{A,A}(\mathbf{R})=0$ and
$a_{i,\alpha}^{D,A}(\mathbf{R})=-a_{i,\alpha}^{A,D}(\mathbf{R})=\frac{1}{M_i}  \int \psi_D(\mathbf{r},\mathbf{R})  \frac{\partial \psi_A(\mathbf{r},\mathbf{R}}
{\partial R_{i,\alpha}}  d\mathbf{r}$  real. We define vector $\mathbf{A}(\mathbf{R})= \{a_{i,\alpha}^{D,A}(\mathbf{R})\}=\{-a_{i,\alpha}^{A,D}(\mathbf{R})\}$.
We also define the matrix elements for $n,m=D$ or $A$,
 $ w^{n,m}(\mathbf{R})=w^{m,n}(\mathbf{R}) =
\sum_i\frac{1}{2M_i} \int \nabla_{\mathbf{R}_i}\psi_n(\mathbf{r},\mathbf{R}).\nabla_{\mathbf{R}_i}\psi_m(\mathbf{r},\mathbf{R}) d\mathbf{r}$.
Then,  the projected kinetic operator $\sum_i \frac{\mathbf{P}_i^2}{2M_i}$ becomes the $2\times2$ matrix of operators
$ \tilde{K} =\left(\begin{array}{ cc}
\mathbf{K}_{DD}     &  \mathbf{K}_{DA} \\
\mathbf{K}_{AD} &  \mathbf{K}_{AA} 
\end{array} \right) $
where
$\mathbf{K}_{DD}= \sum_i \frac{\mathbf{P}_i^2}{2M_i} + \hbar^2  w^{D,D}(\mathbf{R}) $,
$ \mathbf{K}_{AA}= \sum_i \frac{\mathbf{P}_i^2}{2M_i} + \hbar^2  w^{A,A}(\mathbf{R})$,
$\mathbf{K}_{DA}= -\frac{i \hbar }{2}(\mathbf{A}.\mathbf{P}+\mathbf{P}.\mathbf{A}) 
+\frac{\hbar^2}{2} \mathbf{\nabla}. \mathbf{A} +\hbar^2 w^{D,A}(\mathbf{R})=\mathbf{K}_{AD}^{\star}$.
It is convenient to use the base of the Pauli matrices  $\sigma^{x}, \sigma^{y}, \sigma^{z}$
% $\sigma^{x} = \left(\begin{array}{cc} 0 & 1 \\ 1 & 0 \end{array}\right) 
% \qquad \sigma^{y} = \left(\begin{array}{cc} 0 & -i \\ i & 0 \end{array}\right) 
%\qquad \sigma^{z} = \left(\begin{array}{cc} 1 & 0 \\ 0 & -1 \end{array}\right) $
which fulfills the standard commutation relations
$[ \sigma^x,\sigma^y] = 2i \sigma^z$, $[ \sigma^y,\sigma^z] = 2i \sigma^x$, $[ \sigma^z,\sigma^x] = 2i \sigma^y$
so that the fully quantum Hamiltonian appears a collection of quantum nuclei coupled to a quantum spin 
%(representing the electronic degree of freedom in the restricted space $\mathcal{E}(\mathbf{R})$)
\begin{eqnarray}
\tilde{H}&=&\sum_i \frac{\mathbf{P}_i^2}{2M_i}  +\mathcal{V}_{ph}(\mathbf{R}) \nonumber \\
&+& \tilde{\Lambda}(\mathbf{R}) \sigma^x+  \Pi(\mathbf{R}, \mathbf{P})  \sigma^y +  \mathcal{W}(\mathbf{R}) \sigma^z
 \label{hamspin}
 \end{eqnarray}
where $\mathcal{V}_{ph}(\mathbf{R}) = \frac{1}{2} (E_D(\mathbf{R})+E_A(\mathbf{R}))
+\frac{\hbar^2}{2} (w^{D,D}(\mathbf{R})+w^{A,A}(\mathbf{R}))$,
$\mathcal{W}(\mathbf{R})= \frac{1}{2} (E_D(\mathbf{R})-E_A(\mathbf{R}))
+\frac{\hbar^2}{2} (w^{D,D}(\mathbf{R})-w^{A,A}(\mathbf{R}))$,
$\tilde{\Lambda}(\mathbf{R}) = \Lambda(\mathbf{R}) +\frac{\hbar^2}{2} \mathbf{\nabla}. \mathbf{A}(\mathbf{R}) +\hbar^2 w^{D,A}(\mathbf{R})$ and
 $\Pi(\mathbf{R}, \mathbf{P}) = \frac{ \hbar }{2}\left(\mathbf{A}(\mathbf{R}).\mathbf{P}+\mathbf{P}.\mathbf{A}(\mathbf{R}) \right)$.

For going further, it is now convenient to assume  that  1)  The origin of the nuclei coordinates as well as the origin of the energies 
are chosen at the minimum of potential  $\mathcal{V}_{ph}(\mathbf{R})$ which is supposed to be quadratic with 
the elasticity matrix $\overline{\overline{\mathbf{M}}}$:
$\mathcal{V}_{ph}(\mathbf{R}) = \frac{1}{2} \mathbf{R}.\overline{\overline{\mathbf{M}}}. \mathbf{R} $.
2) We assume a linear behavior for  the spin coefficients  of $\sigma^z$ (charge coupling), $\mathcal{W}(\mathbf{R}) =  \mathcal{W}(\mathbf{0}) +  \mathbf{\nabla}.\mathcal{W}(\mathbf{0}) .\mathbf{R}$,
of $\sigma^x$ (covalent coupling) $\tilde{\Lambda}(\mathbf{R}) = \tilde{\Lambda}(\mathbf{0}) + \mathbf{\nabla}. \tilde{\Lambda}(\mathbf{0}).\mathbf{R}$ and
of $\sigma^y$ $\Pi(\mathbf{R}, \mathbf{P})= \hbar \mathbf{A}(\mathbf{0}).\mathbf{P}$ (see. \cite{footnote}).
Considering the overlap integrals $a_{i,\alpha}^{D,A}(\mathbf{R})$ may be small with the LCAO base, $\mathbf{A}(\mathbf{0})=0$
for simplicity (though it would not be a big deal to conserve the coupling with $\sigma^y$). 
Then, it is convenient to use the base of normal modes obtained by diagonalization of matrix $\overline{\overline{\mathbf{M}}}$.

Hamiltonian (\ref{hamspin}) becomes those of a single quantum spin $1/2$  submitted to a field  $(\epsilon_x,0,\epsilon_z)$ and
coupled a collection of normal modes $n$ (harmonic oscillators with unit mass  and frequency $\omega_n$) 
by  constants $k_n^x,0,k_n^z$.
\begin{equation}
\tilde{H}= \sum_n \left( \frac{1}{2} \left(p_n^2 + \omega_n^2 q_n^2 \right) +k_n^x q_n \sigma^x +k_n^z q_n \sigma^z \right) + \epsilon_x \sigma^x + \epsilon_z \sigma^z
\label{hamtot4}
\end{equation}

Finally, we use a mean field  approximation where the coupling terms $q_n \sigma^z$  (and similarly for  $q_n \sigma^x$) are replaced by
$q_n \bar{\sigma}^z +\bar{q}_n \sigma^z -\bar{q}_n  \bar{\sigma}^z $ then neglecting  the fluctuation operators $(q_n -\bar{q}_n)( \sigma^z-\bar{\sigma}^z)$
\cite{footnote1}.  We argue the validity of this approximation because the nuclei displacements generated by the molecular reorganization during ET are generally
much larger than their  zero point quantum fluctuations (especially in biomolecules). 
Then, it  comes out after some calculations that the equations of the time evolution of the observables
 $\{\bar{p}_n,\bar{q}_n\}$ and $\bar{\sigma}^x,\bar{\sigma}^y,\bar{\sigma}^z$ form a closed set (that is do not involve any other observables) which 
are identical to the dynamical equations of  a classical  Hamiltonian  (\ref{hamtot4}) where
 the quantum operators $\{p_n,q_n\}$ and $\sigma^x,\sigma^y,\sigma^z$ are just replaced by their real observables $\{\bar{p}_n,\bar{q}_n\}$ and $\bar{\sigma}^x,\bar{\sigma}^y,\bar{\sigma}^z$. It turns out that  quantum spin $\varphi_D(t) | \uparrow > + \varphi_A(t) |\downarrow>$ represented in the eigenbase of $\sigma^z$ (Donor Acceptor)  
turns out to be submitted to the time dependant classical field
with components $(\epsilon_x+ \sum_n k_n^x \bar{q}_n(t),0,\epsilon_z +\sum_n k_n^z \bar{q}_n(t))$
while the classical oscillators $n$ are submitted to external time dependant forces $f_n(t)=k_n^x \bar{\sigma}^x(t)+k_n^z \bar{\sigma}^z(t)$.

Actually, the motion $\bar{q}_n(t)$ of each oscillator may be explicitly calculated as the sum of a function of the external force $f_n(t)$
and a solution of the free oscillator  chosen randomly according to the Boltzman statistics as done in \cite{AK03,Aub07} so that the nuclei variables
be eliminated from the dynamical equation describing the electron dynamics. The spin (or electronic) dynamics is then described by
\begin{eqnarray}
 i \hbar \dot{\varphi}_D &=& \frac{\partial H_{eff}}{\partial \varphi_D^{\star}} + \zeta^z(t) \varphi_D + \zeta^x(t) \varphi_A \nonumber \\
&+&\left(  \int_0^t \left(\Gamma_{zz}(t-\tau) \dot{Z}(\tau) + \Gamma_{xz}(t-\tau) \dot{X}(\tau)\right) d\tau\right) \varphi_D \nonumber \\
&+& \left(  \int_0^t \left(\Gamma_{xz}(t-\tau) \dot{Z}(\tau) + \Gamma_{xx}(t-\tau) \dot{X}(\tau)\right) d\tau\right) \varphi_A \nonumber \\
 i \hbar \dot{\varphi}_A &=& \frac{\partial H_{eff}}{\partial \varphi_A^{\star}} - \zeta^z(t) \varphi_A + \zeta^x(t) \varphi_D \nonumber \\
&-&\left(  \int_0^t \left(\Gamma_{zz}(t-\tau) \dot{Z}(\tau) + \Gamma_{xz}(t-\tau) \dot{X}(\tau)\right) d\tau\right) \varphi_A \nonumber \\
&+& \left(  \int_0^t \left(\Gamma_{xz}(t-\tau) \dot{Z}(\tau) + \Gamma_{xx}(t-\tau) \dot{X}(\tau)\right) d\tau\right) \varphi_D \nonumber \\
\label{neweq}
\end{eqnarray}
where the effective Hamiltonian is
\begin{eqnarray}
H_{eff} (\varphi_D,\varphi_A)&=& \epsilon_x X + \epsilon_z Z \nonumber\\ - \frac{1}{2} \Gamma_{xx}(0) X^2 &-& \Gamma_{xz}(0) XZ - \frac{1}{2}  \Gamma_{zz}(0) Z^2 
 \label{generham}
\end{eqnarray}
and where $Z=\bar{\sigma}^z=  |\varphi_D|^2-|\varphi_A|^2$ and $X=\bar{\sigma}^x=\varphi_D^{\star}\varphi_A+\varphi_D\varphi_A^{\star}$.
The memory functions are defined as
$\Gamma_{xx}(t) = \sum_n \frac{(k_n^x)^2}{\omega_n^2} \cos \omega_n t = \int \tilde{\Gamma}_{xx}(\omega)   \cos \omega t ~d\omega$,
$\Gamma_{zz}(t) =\sum_n \frac{(k_n^z)^2}{\omega_n^2} \cos \omega_n t = \int \tilde{\Gamma}_{zz}(\omega)  \cos \omega t ~d\omega $,
$\Gamma_{xz}(t) = \Gamma_{zx}(t)=\sum_n \frac{k_n^z k_n^x}{\omega_n^2} \cos \omega_n t$
and fulfill $0 \leq \Gamma_{zz}(0)$, $0 \leq \Gamma_{xx}(0)$ and $ \Gamma_{zx}^2(0) \leq  \Gamma_{zz}(0) \Gamma_{xx}(0)$.
Assuming a large system with many classical oscillators generates a classical Langevin bath \cite{footnote2}. Thus, these functions may be assumed to be smooth and to vanish at large time. 
but with   Fourier transforms which vanish for $\omega>\omega_c$ where $\omega_c$ is the maximum phonon frequency.
It can be shown indeed  that in the absence of random forces  the effective energy $H_{eff}$ necessarily decay
(essentially due to  non adiabaticity) \cite{Aub07} but the damping  is  frequency dependant. Actually when the characteristic frequency 
of the electronic dynamics goes beyond $\omega_c$, the effect of damping disappears. The non adiabatic effects 
thus disappear so that the system obeys again the BO approximation. 

This effective Hamiltonian (\ref{generham}) may be obtained as the energy minimum of the whole system
at fixed electronic amplitudes $\varphi_D$,$\varphi_A$ and with respect to all the nuclei coordinates.
A similar one was introduced phenomenologically in \cite{AK03} 
 without microscopic justifications but  the form which was used, was not correct because we  omitted covalent terms
  and artificially introduced extra nonlinear capacitive terms. 
  These terms actually do not exist because already taken into account in the coefficients of the linear quantum electronic Hamiltonian $H_{el}$.
%, \nonumber \\
%&=& \int \tilde{\Gamma}_{xz}(\omega)  \cos \omega t ~d\omega 
%\label{gamdef}
%\end{eqnarray}
Temperature appears  in eqs.(\ref{neweq}) through the gaussian random forces  with correlations fulfilling the Langevin conditions
$<\zeta^z(t)\zeta^z(t+\tau)>_t = k_B T  \Gamma_{zz}(\tau)$,% \nonumber \\
$<\zeta^x(t)\zeta^x(t+\tau)>_t = k_B T  \Gamma_{xx}(\tau)$,% \nonumber \\
$<\zeta^x(t)\zeta^z(t+\tau)>_t = k_B T  \Gamma_{xz}(\tau)$ % \label{Lang2}

\begin{figure} 
    \centering
      \includegraphics[width=0.23 \textwidth]{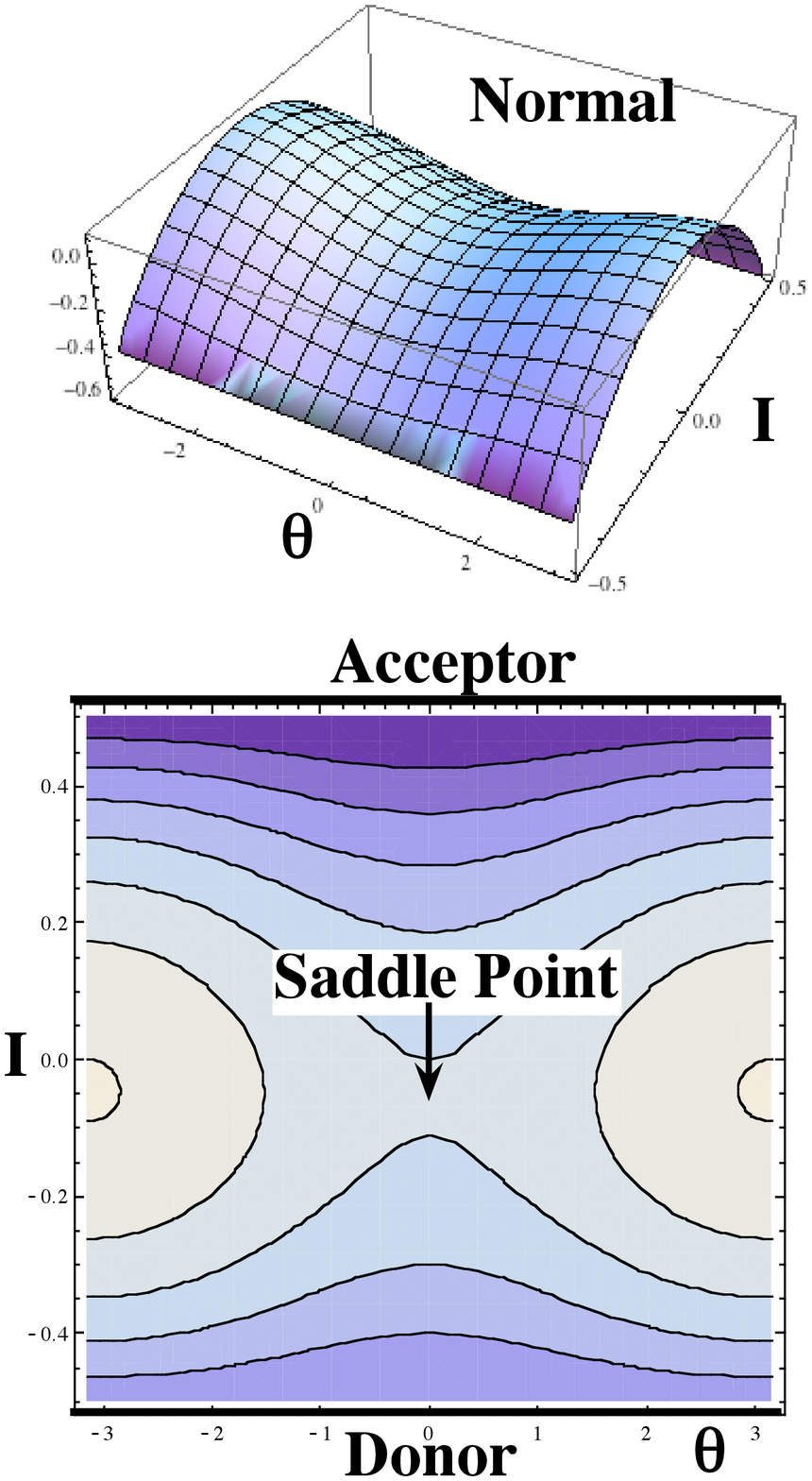}
        \includegraphics[width=0.23 \textwidth]{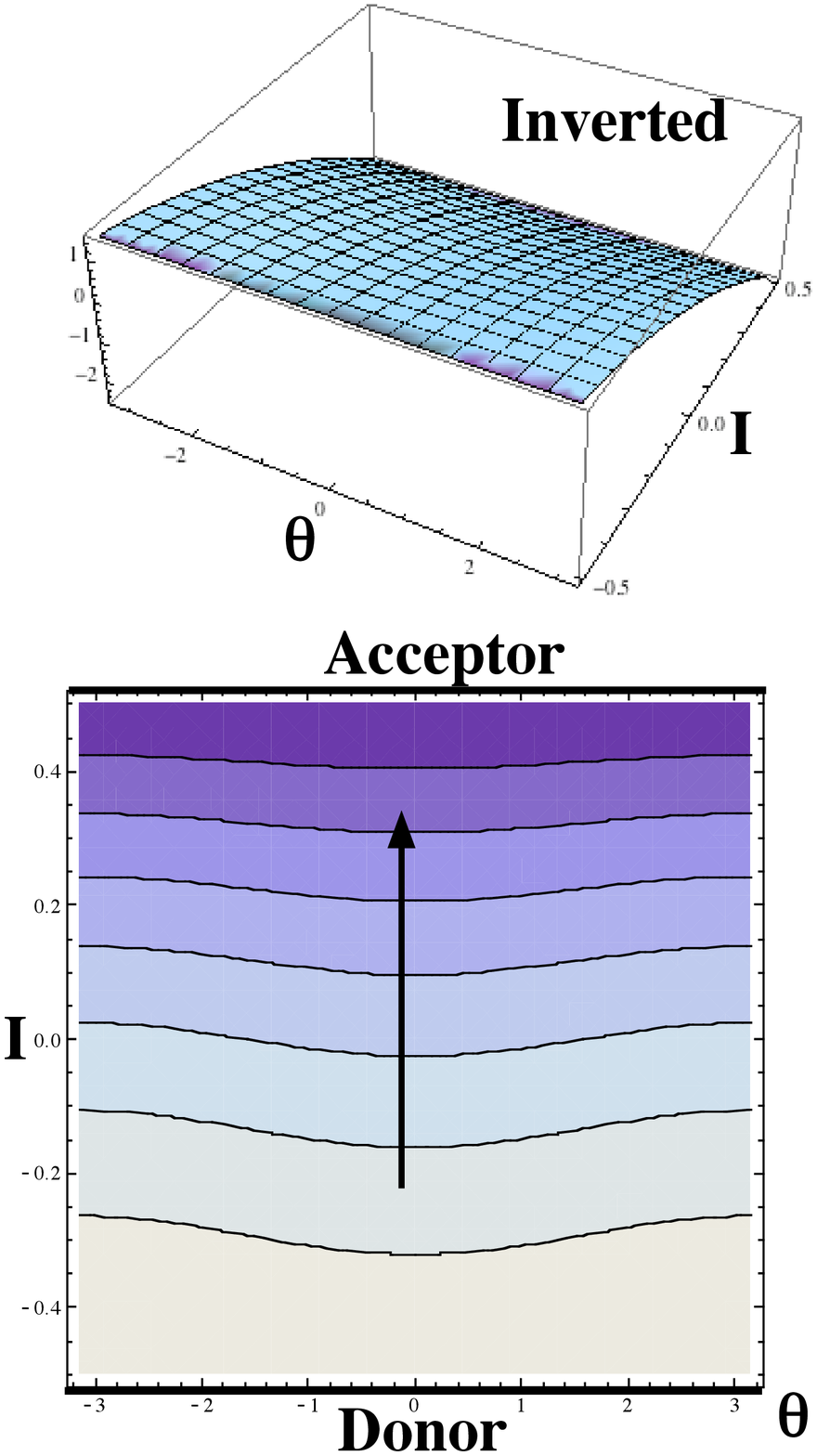}
    \caption{Some 3D plot and below their corresponding contour plots for the energy lanscape $H_{eff}(I,\theta)$ on the sphere $I,\theta$
    (represented with the Mercator projection where the poles are single points  appearing as the thick lines)
    in cases with no covalent interactions ($\Gamma_{xx}(0)=\Gamma_{xz}(0)=0$) and  $\Gamma_{zz}(0)=1.$\\
On the left side : $\epsilon_z=0.1$, $\epsilon_x=-0.1$ corresponds to a normal regime in the standard theory (see fig.\ref{fig1})
and on the right side $\epsilon_z=1$, $\epsilon_x=-0.1$ to an inverted regime. }
 \label{figsur1}
\end{figure}

\begin{figure} 
    \centering
      \includegraphics[width=0.23 \textwidth]{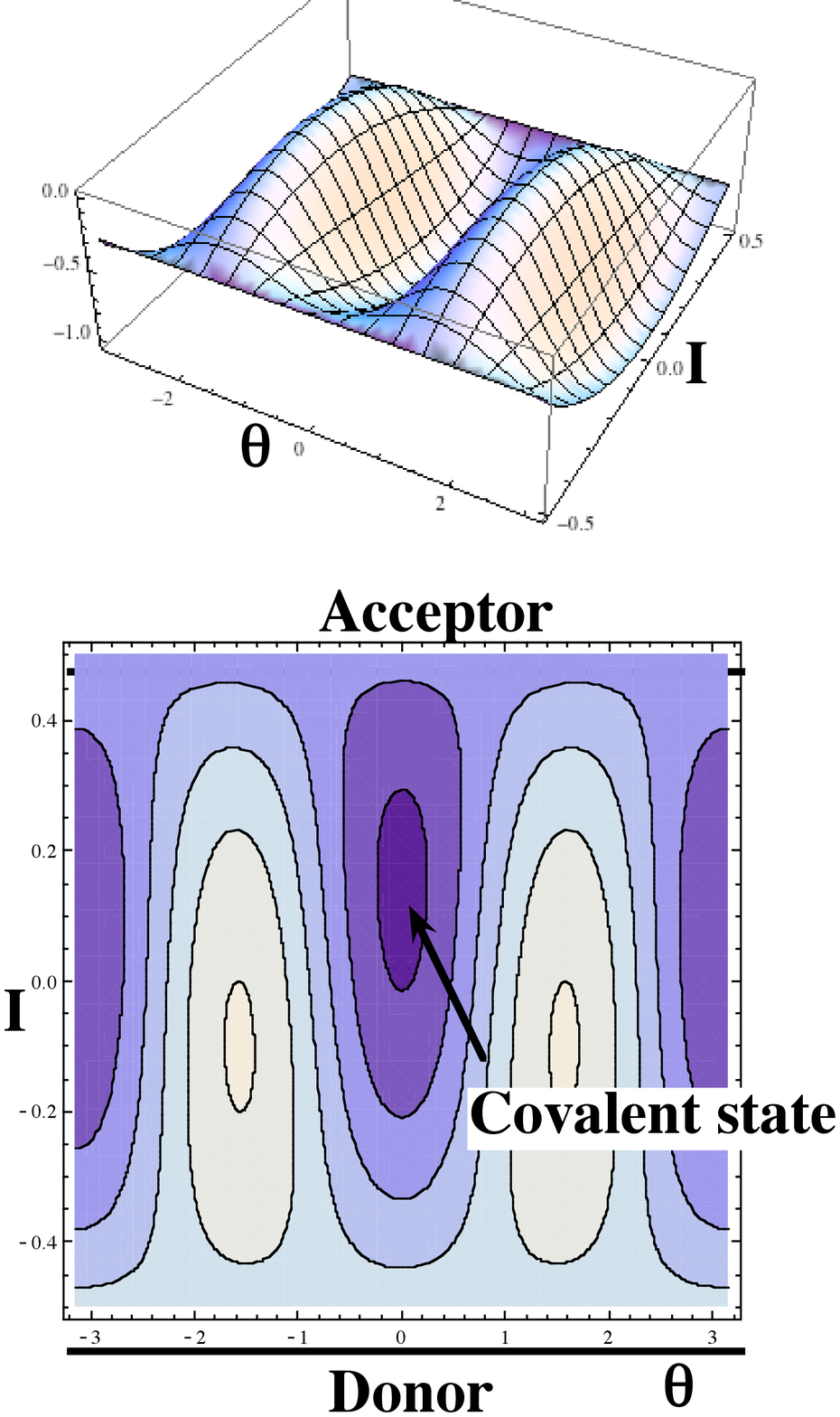}
        \includegraphics[width=0.23 \textwidth]{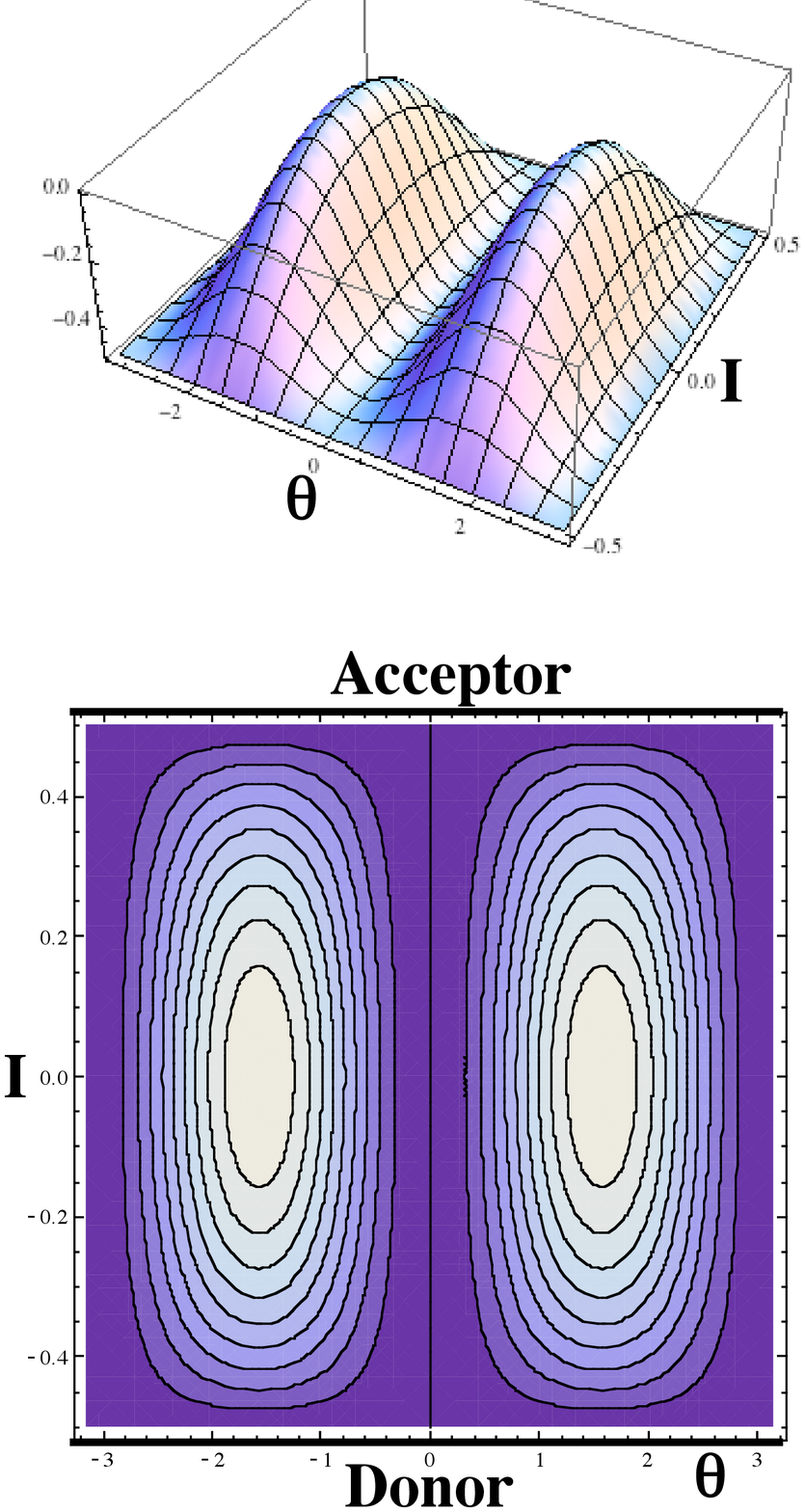}
    \caption{Same as fig.\ref{figsur1} but for some examples with covalent interactions. 
   On the left  side $\epsilon_z=0.2,\epsilon_x=-0.01,\Gamma_{zz}(0)=1,\Gamma_{xx}(0)=2.$ and $\Gamma_{xz}(0)=-0.1$.
correspond to a situation where the final state is a covalent state.
On the right side : $\epsilon_z=\epsilon_x=0$, $\Gamma_{xx}(0)=\Gamma_{zz}(0)=1.$ and $\Gamma_{xz}(0)=0$) 
  correspond to an ideally isoenergetic and barrierless situation.}
 \label{figsur2}
\end{figure}

Defining new conjugate variables $I_D,\theta_D,I_A,\theta_A$ by $\varphi_D=\sqrt{I_D}e^{-i\theta_D}$ and $\varphi_A=\sqrt{I_A}e^{-i\theta_A}$, and
next the conjugate variables $I=(I_A-I_D)/2= -Z/2$ and $\theta=\theta_A-\theta_D$,
this effective Hamiltonian becomes only a function $-\frac{1}{2}\leq I \leq \frac{1}{2},\theta$ mod $2\pi$
on a sphere with poles $I=\pm \frac{1}{2}$ and where
$\theta$ corresponds to the longitude and $\phi$ defined as $\sin \phi = 2*I$ to the latitude. 
Then  $H_{eff}=  -2 \epsilon_z I+ \epsilon_x \sqrt{1-4I^2}\cos \theta-2 \Gamma_{zz}(0) I^2+
2 \Gamma_{xz}(0)  I  \sqrt{1-4I^2}\cos \theta  - \frac{1}{2} \Gamma_{xx}(0) (1-4I^2) \cos^2 \theta $
represents the true energy surface for ET. At zero temperature (0K) and assuming
the damping terms vanishes, the dynamical equations would be those of an integrable Hamiltonian system on a sphere.
All trajectories would be periodic on closed orbits defined by a constant  energy $H_{eff}(I,\theta)$.
Figs.\ref{figsur1} and \ref{figsur2} show both 3D and contour plots for several examples.

We assume that the (initial) transfer integral $\epsilon_x$ is small as in the standard theory and that
our system is  initially at the Donor pole $I=-1/2$. The Acceptor pole
corresponds to $I=+1/2$. When there are no covalent interactions ($\Gamma_{xx}(0)=\Gamma_{xz}(0)=0$), our theory
is nothing but a different representation of the standard Marcus theory except that we now have
explicit dynamical equations which intrinsically describe  the diabatic transitions through the damping terms.
without needing the Landau-Zeener model. The two Marcus energy surfaces are the paraboloids obtained from Hamiltonian (\ref{hamtot4})
where $\sigma^x=0,p_n=0$ and  $\sigma^z=+1$ (electron on Donor) or   $\sigma^z=-1$ (electron on Acceptor)
Then the reaction energy is $\Delta G_0=2  \epsilon_z$ assumed to be positive, the barrier energy $\Delta G^{\star}=\frac{(\epsilon_z - \Gamma_{zz}(0))^2}{2 \Gamma_{zz}(0)}$
and the electronic excitation energy $E_{el}= 2 (\Gamma_{zz}(0)-\epsilon_z)$.

In the normal regime, when $E_{el}>0$, the energy surface (see fig.\ref{figsur1} left) exhibits a saddle point between donor and acceptor
which corresponds to the barrier between the donor and acceptor state. At the inversion point where $ E_{el}=0$, the saddle point 
and the two maxima on the sphere of $H_{eff}(I,\theta)$ merge with the  minimum near the pole $I=-1/2$ which thus become a single maximum
so that  in the inverted regime $E_{el}<0$, there is no more energy barrier (see fig.\ref{figsur1} right).
In both regimes, the two poles on the sphere Donor and Acceptor are surrounded by periodic and stable orbits with  frequency $\omega_{el}$
obtained by linearizing the equations (\ref{neweq}) which turns out to be related to $|E_{el}|=\hbar \omega_{el}$. 
When this electronic frequency $\omega_{el}$  is beyond the phonon spectrum that is $\omega_{el}>>\omega_c$,  the damping terms have no 
dissipative effect in eqs.\ref{neweq} so that the poles are really stable at 0K. Then, the thermal random forces  in eq.\ref{neweq}
are necessary to bring the electronic levels near resonance where the damping terms become efficient in eqs.\ref{neweq}.  
We have already shown in \cite{Aub07} that reaching this resonance is equivalent to reach the intersection between the two paraboloids
so that we recover an Arrhenius law.
On contrary, in the inverted regime but near the inversion point  where  $\omega_{el} < \omega_c$, ET spontaneously occurs at 0K (and very fast)
since the phonon bath can absorb efficiently the reaction energy.

When  $\Gamma_{zz}(0) << \Gamma_{xx}(0) $ and $\epsilon_z <\Gamma_{xx}(0) $ (then $\Gamma_{xx}(0)>> |\Gamma_{xz}(0)|$),
there are two energy minima on the sphere (see fig.\ref{figsur2} left) corresponding both to covalent bonds
where $\theta=0$mod $\pi$  and $I=I_c\approx \frac{\epsilon_z}{2\Gamma_{xx}(0)}$ with $|I_c|<1/2$ and  two maxima  on  $\theta=\pi/2$ mod $\pi$.
Actually, only the lowest minimum is physically acceptable for the covalent bond  \cite{footnote3}. The poles $I=\pm 1/2$ are unstable because they belong to large amplitude time periodic orbits with a low frequency in the range of phonon frequencies. These trajectories are dissipative and start to converge toward  the minimum 
energy solution which is the covalent bond. This is the situation of free radicals which spontaneouly bind without activation energy. 

The most interesting situation is obtained in the intermediate case, when  both charge and covalent interactions are present.
Fig.\ref{figsur2} shows the ideal case obtained for well chosen parameters where $\epsilon_x=\epsilon_z=0$, $\Gamma_{xx}(0)=\Gamma_{zz}(0)$,
$\Gamma_{xz}(0)$. Then $H_{eff}(I,\theta)$ is minimum along two degenerate paths $\theta=0$ or $\pi/2$ which is quite similar to those
of a dimer model with Targeted Energy Transfer \cite{AKMT01}. Again because of \cite{footnote3}, only one of these paths physically exist.
 Actually for model parameters near but not equal to their ideal values,  the energy profile between Donor and Acceptor is rather flat 
with small reaction energy for ET positive or negative.  When the reaction energy is positive without energy barrier
ET occurs spontaneously and very fast at 0K because the phonon bath is dissipative. ET may be reversed when the reaction energy is negative.
If there is a small energy barrier, ET is nevertheless ultrafast but requires a small temperature.

In summary, we have shown that using the complex electronic amplitudes as reaction coordinates (instead of the nuclei coordinates) allows
one to treat both charge and covalent interactions. In the limit where only charge interactions are present, 
we recover the standard redox theory of ET (apart  the Landau-Zeener effect concerning
the prefactor of the Arrhenius law). With only the covalent interactions, we can model the covalent binding of free radicals.
The most interesting result is  the possible existence in the intermediate regime of almost barrierless and isoenergetic chemical paths 
for a two state (or dimer) model as needed for understanding biochemistry. 

In further publications, we shall extend this formalism to three states models and in particular 
reconstruct  a physically correct model for ultrafast electron transfer reproducing the same features as in \cite{AK03,Aub07}.
Variations/extensions  of this trimer model will be proposed for ultrafast charge transport (or excitonic) in conducting polymers or along selected paths in proteins,
 for highly energetic reactions with excitonic storage of the reaction energy, for  biomotors (where motion is obtained through molecular reorganisation) etc...

Our enzymatic models require fine tuning of their parameters so that their function can be either blocked, 
activated or even reversed by changes of the environment involving the solvant (pH, ions concentration)
the binding/unbinding  of cofactors, changes of conformation etc...  then allowing regulation and logics. 
 We believe that complex enzymes involve transitions within a large set of  electronic states organized  in such a way they operate
as logical processing units.

I acknowledge George Kopidakis and Jos\'e Teixeira for valuable discussions and Laboratoire L\'eon Brillouin for its hospitality.


\begin{thebibliography}{99}
\bibitem{footnote} The assumption that the wave functions defining the base of the invariant subspace 
$\psi_D(\mathbf{r},\mathbf{R})$ and $\psi_A(\mathbf{r},\mathbf{R})$ and their overlaps are smoothly dependant on $\mathbf{R}$,
may be reasonably assumed for the LCAO base but not for other bases such as those of electronic eigenstates.
\bibitem{footnote1} If $\Psi(0)$ is the wave function at time $0$
and $A(t) = e^{\frac{i}{\hbar} \tilde{H}t} A e^{-\frac{i}{\hbar} \tilde{H}t}$, the observable of operator $A$ at time $t$ is $\bar{A}(t)= <\Psi(0)|A(t)|\Psi(0)>$. Then $i \hbar \dot{\bar{A}}(t)=\overline{[\tilde{H},A]}$.
\bibitem{footnote2} We may also assume only few damped modes thermalized by white random forces instead of a phonon continuum. Such an assumption may be 
convenient for applications.
\bibitem{footnote3}  $\Lambda(\mathbf{R})$ should have a constant sign and exponentially vanish at large distance Donor-Acceptor.
However, because of the linear approximation on $\Lambda(\mathbf{R})$ in (\ref{hamspin}), it may  change sign
instead of vanishing. Thus potential $H_{eff}(I,\theta)$ is well-described only on the part of the sphere where $\Lambda(\mathbf{R})$
keep the same sign as $\Lambda(\mathbf{0})$. Thus, one of its two minima is an artefact which does not correspond physically to a covalent bond.
\end{thebibliography}
\end{document}